# Modern Cybersecurity Solution using Supervised Machine Learning

Mustafa Sakhai
msakhai@agh.edu.pl
AGH University of Science and Technology
Kraków, Poland

Maciej Wielgosz
wielgosz@agh.edu.pl
AGH University of Science and Technology
Kraków, Poland

## Abstract:

Cybersecurity is essential, and attacks are rapidly growing and getting more challenging to detect. The traditional Firewall and Intrusion Detection system, even though it is widely used and recommended but it fails to detect new attacks, zero-day attacks, and traffic patterns that do not match with any configured rules. Therefore, Machine Learning (ML) can be an efficient and cost-reduced solution in cybersecurity.

We used Netflow datasets to extract features after applying data analysis. Then, a selection process has been applied to compare these features with one another. Our experiments focus on how efficient machine learning algorithms can detect Bot traffic, Malware traffic, and background traffic. We managed to get 0.903 precision value from a dataset that has 6.5% Bot flows, 1.57% Normal flows, 0.18% Command&Control (C&C) flows, and 91.7% background flows, from 2,753,884 total flows. The results show low false-negative with few false-positive detections.

# Introduction:

In today's world, we perform most of our daily activities over the Internet, from online banking, shopping, meetings, sharing and storing data to online doctor appointments. That makes security need is essential and crucial to secure endpoints, edge networks, data centres, and applications that make the Internet that we know today [1]. There are different security solutions in the market, either commercial or open-source solutions, that can help to detect and monitor threats.

The Intrusion Detection System (IDS) solution is commonly used to collect data from ongoing attacks, so the collected data can be used later on to tune security configurations and policies. Usually, a honeypot is used with an IDS, a system with known manageable vulnerabilities that helps understand attacks and attackers methodology [2]. There is an Intrusion Prevention System (IPS) similar to Intrusion Detection System (IDS). However, IPS has rule-based policies that can take action on traffic in real-time, Block, Monitor, or Allow a particular connection [3]. A well-known IDS/IPS solution is Snort, a widely-used Open Source Intrusion Prevention System (IPS). Snort IPS uses a pre-defined series of rules that help define network malicious activity and generates alerts for users [10]. Most of the existing solutions are only signature-based intrusion systems designed to detect known attack patterns. It will require extensive research of new attacks and regular updating of the rule-base signatures [5]. There are well-known attacks like IRC (Internet Relay Chat); protocol-based botnet for control and command infected Internet computers [6]. SPAM; is used to describe similar abuses in other media and mediums [7]. Click Fraud (CF); attacks employ large distributed botnets, deceptive publisher pages, malware infection, and fake conversion "chaff" to cloak fraudulent activity [8]. Furthermore, Port Scan (PS); used to scan and check open 'listening' ports of a host, a server, or a node to prepare for an attack or use a vulnerability [9]. Even though security researchers have done much research in the security field, all these attacks are still dangerous and can cause a severe breach.

Anomaly detection systems detect malicious activities by identifying any deviation from standard traffic patterns [9]. It also does not require the amount of time and effort used in network intrusion detection systems and host-based intrusion detection systems to update rule-based signatures because there are no signatures. It requires time to train and test a model with enough data to recognise and record the normal activities than to build a Normal pattern. Some studies describe anomaly detection in cybersecurity using machine learning algorithms [9][11][15].

There are widely used terms in machine learning and security: False Positive and False Negative, and these terms reflect how a model is precise in detecting anomalies [9]. This paper presents a comprehensive experiment about supervised machine learning algorithms for anomaly detection. We used the CTU-13 dataset [16], and we have tested different scenarios and parameters. The contribution of this paper is to describe how effective machine learning (ML) is by tuning specific parameters, and in the end, we conclude where this model can fit the best.

This paper is structured as the following; In Section 2, we present some of the related work. In Section 3, we explain the dataset, Section 4, we explain the methodology and feature selection, Section 4, devoted to the experimental analysis using Machine Learning (ML) algorithms. Finally, conclusions are given in Section 6.

## 1) Contribution

We have run experiments to test a wide range of parameters in the Supervised Machine Learning algorithm, which has not been tested such deeply before. We have successfully built a table to identify the optimal window width and stride values that can be used for Normal traffic, bot traffic, and malware traffic classification. We could test and identify where each model can be applied for better malware protection with low false-negative and false-positive detectors.

## 2) Literature Review

In [11], the authors performed an extensive review of the literature and original experiments on real, large enterprises and network traffic. The authors have categorised Machine Learning as Shallow Learning and Deep Learning, each being categorised as supervised and unsupervised algorithms.
In [13], phishing was explored using C4.5, decision tree, and other approaches which include Random Forest, Support Vector Machine and Naïve Bayes, (PILFER), "Phishing Identification by Learning on Features of Email Received", was developed as an anti-phishing method and then investigated on a set of 860 phishing and 695 ham cases. The results were different features for recognising instances as phishing or ham, i.e. IP URLs, time of space, HTML messages, number of associations inside the email, JavaScript and others. Hence, the authors explained that PILFER could improve the clustering of messages by joining all ten features found in the classifier beside " Spam filter output " [14]. However, Logistic Regression was explored as part of that research.
In [14], the authors made a comparison between phishing detection models content and features; The experimental section demonstrates that the knowledge-based approach presented by Ridor and eDRI algorithms seem to be appropriate to combat phishing for two reasons
1) The classification accuracy of the models derived are highly competitive
2) The models contain easy to understand knowledge by novice users so they can utilise them in making decisions
Decision trees Bayes Net and SVM achieved good detection rates.
In [15], the authors applied four different deep learning models, analyses namely Convolutional Neural Network (CNN), Long Short- Term Memory (LSTM), hybrid CNN-LSTM, and Multi-layer Perception (MLP) for bot detection and simulation using the CTU-13 dataset, the studies show 100% precision via using CNN model for Scenario 3 which contains only 24 botnet flows which are a very small portion of the total number of flows in Scenario 3 which is 27457 flows, that's made us concern if the model will perform well with other datasets that have more malicious traffic. The same authors showed that using CNN-LSTM and MLP gives 100% of Accuracy, Sensitivity, Specificity, Precision, and F1 Score. We will be using the same dataset and applying supervised machine learning Logistic Regression.

The implicit cost of misclassification in the cybersecurity domain is a serious problem. False positives in malware classification and intrusion detection annoy security operators and hinder remediation in case of actual infection [11]. In our malware analysis, we consider the approach in [9]. We have picked up the supervised machine learning algorithm "Logistic Regression" and applied a detailed analysis and comparison between different parameters which were trained and tested on the same dataset in depth.

## 3) Dataset

To evaluate the supervised machine learning model, we use the CTU-13 dataset, a botnet traffic dataset created in the Czech Republic at CTU University in 2011 [16]. The CTU-13 dataset has a unique combination in one dataset; normal flows, bot flows, background flows, malicious flows. The distinctive characteristic of the CTU-13 dataset is that we manually analysed and label each scenario. The labelling process was done inside the NetFlows files. Table 4 shows the relationship between the number of labels for the Background, Botnet, C&C Channels and Normal on each scenario [16].

The CTU-13 dataset is a dataset in which each traffic flow is labelled as attack, benign, or background traffic. The labelling is performed as follows: attack labels are assigned to all traffic flows between infected systems (bots) and victim systems (attack targets), benign labels are assigned to all other traffic flows from or to the victims, and background labels are assigned to traffic flows not affecting the victims [15].

The CTU-13 dataset has the following attributes obtained by NetFlow;

| Attribute | Description |
|---|---|
| StartTime | The start time when the flow is initiated |
| Dur | Duration of how long the flow is active |
| Proto | Protocol (TCP, UDP, etc.) |
| SrcAddr | The source IP address |
| Sport | The source port |
| Dir | Direction of communication |
| DstAddr | The destination address |
| Dport | The destination port |
| State | The state of a protocol |
| sTos | The source Type of Service |
| dTos | The destination Type of Service |
| TotPkts | The total number of packets exchanged |
| TotBytes | The total bytes exchanged |
| SrcBytes | The number of bytes sent by the source |
| Label | The label assigned to this Netflow |

Table (1): CTU-13 dataset attributes

There are five features extracted [9];

| | |
|---|---|
| | The sum |
| | The mean |
| Extracted Features | The standard deviation |
| | The maximum |
| | The median |

Table (2): The extracted features from the CTU-13 dataset

## 4) Methodology

After the literature review in section 2, we start seeking research that deep dive in using Machine Learning to classify Bot traffic and Malicious traffic. We were looking for models that can be applied to different security scenarios based on a specific business need, like a model with high precision, among F1-score and Recall values that can be applied in high sensitive solutions, or a model with high Recall to capture as much event as possible regardless of the precision of the model, or a model that would combine both acceptable Precision and Recall detection. In [9], authors have tested different Machine Learning (ML) algorithms. However, we could extend the research and deep dive into testing a new and wide range of parameters. We ran our experiments in two parts; in a powerful physical server, and other experiments in the Cloud using Amazon Web Services (AWS) instances. Figure (1) presents the tested values using a Supervised Machine Learning algorithm to obtain different values of Precision, Recall, and F1-score values.

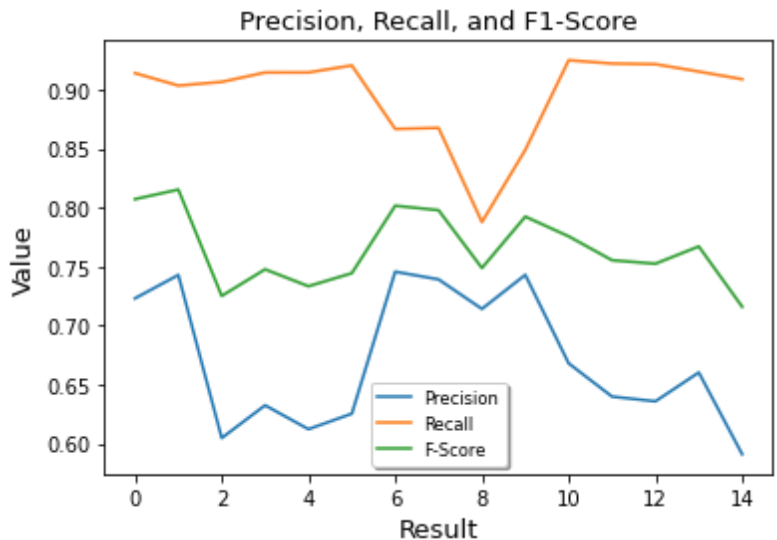

Figure (1): Precision, Recall, and F1-score values

**4.1: Feature Selection**
This step is required to select features from the extracted ones. It involves the use of feature selection techniques to reduce the dimension of the input training matrix. Filtering features through Pearson Correlation, wrapper methods using Backward Feature Elimination, embedded methods within the Random Forest Classifier, Principal Component Analysis (PCA), and the t-distributed Stochastic Neighbour Embedding (t-SNE) are the different techniques used for this purpose.

**4.2: Botnet Detection**
We tested our model to see if it can successfully detect botnet traffic from the CTU-13 dataset [16]. Bot traffic software-generated traffic serves a specific purpose, and bot traffic is almost everywhere. It is crucial to identify which bot traffic is benign and which is malicious. The overall performance of botnet detection is determined from the F1 score.

## 5) Experiments and results

### 5.1: Experiment's environment
We used a Physical desktop server machine with a processor Intel Ci9, RAM 64 GB, 1.5 TB of hard drive, GPU RTX GeForce 3080, and 4 AWS EC2 instances type x1e.32xlarge deployed in the Cloud for high performance and time reduction of experiments.

### 5.2: Logistic Regression
A supervised learning model, "Logistic regression", is used as a way for binary classification. The word itself is obtained from Statistics. At the core of the method, it uses logistic functions, a valuable sigmoid curve for various fields, including neural networks. Logistic regression models the probability for classification problems with two possible outcomes. Logistic regression can be used to identify network traffic as malicious or not.

We got inspired by the research in [9], and we decided to proceed further and train and test the logistic regression model from all aspects.

We expect this paper to be a good reference for applying supervised Machine Learning Logistic Regression in Anomaly Detection when processing and dealing with a large amount of data. Table (3) shows how precision, recall, and F1-score change based on the window width and stride values.

|    | Width(seconds) | Stride(seconds) | Precision | Recall | F1-score |
|----|----------------|-----------------|-----------|--------|----------|
| 1  | 60             | 60              | 0.723     | 0.914  | 0.807    |
| 2  | 90             | 15              | 0.742     | 0.903  | **0.815** |
| 3  | 90             | 75              | 0.604     | 0.906  | 0.725    |
| 4  | 90             | 30              | 0.632     | 0.914  | 0.747    |
| 5  | 90             | 25              | 0.612     | 0.914  | 0.733    |
| 6  | 90             | 90              | 0.625     | 0.920  | 0.744    |
| 7  | 180            | 30              | **0.745** | 0.866  | 0.801    |
| 8  | 180            | 120             | 0.739     | 0.867  | 0.797    |
| 9  | 165            | 105             | 0.737     | 0.864  | 0.795    |
| 10 | 189            | 129             | 0.739     | 0.877  | 0.801    |
| 11 | 480            | 180             | 0.714     | 0.787  | 0.748    |
| 12 | 600            | 15              | 0.743     | 0.848  | 0.792    |

| 13 | 135 | 15 | 0.668 | **0.924** | 0.775 |
| 14 | 75  | 15 | 0.639 | 0.922 | 0.755 |
| 15 | 75  | 15 | 0.635 | 0.921 | 0.752 |
| 16 | 150 | 15 | 0.660 | 0.915 | 0.767 |
| 17 | 60  | 60 | 0.591 | 0.908 | 0.716 |

Table (3): Calculating Precision, Recall, and F1-score using different width and stride window values

In Table (4), we have selected a value width = 1 minute and stride = 1 minute, and we ran the same experiment (4) times to confirm the predicted values are real and connected.

|   | Train | | | | Test | | |
|---|---|---|---|---|---|---|---|
|   | Precision | Recall | F1 Score | Time (hrs) | Precision | Recall | F1 score |
| 1 | 0.602 | 0.925 | 0.729 | 7.5 | 0.598 | 0.921 | 0.725 |
| 2 | 0.573 | 0.914 | 0.704 | 7.8 | 0.571 | 0.913 | 0.702 |
| 3 | 0.583 | 0.915 | 0.712 | 7.4 | 0.586 | 0.914 | 0.714 |
| 4 | 0.590 | 0.911 | 0.716 | 7.6 | 0.589 | 0.910 | 0.714 |

Table (4): Running width and stride that equal to 1 minute four times to confirm the values

## 5.3: Better prediction using Logistic Regression

We have carefully tested and applied different changes to the data to understand where logistic regression can best be applied. Therefore, we have tested different data sets from the CTU-13 dataset [16]. In Table (5), we can understand how different labels are distributed for each dataset scenario.

| Scenario | Total Flows | Botnet Flows | Normal Flows | C&C Flows (Command and Control) | Background Flows |
|---|---|---|---|---|---|
| 1 | 2,824,636 | 1.41% | 1.07% | 0.03% | 97.47% |
| 2 | 1,808,122 | 1.04% | 0.5% | 0.11% | 98.33% |
| 3 | 4,710,638 | 0.56% | 2.48% | 0.001% | 96.94% |
| 4 | 1,121,076 | 0.15% | 2.25% | 0.004% | 97.58% |
| 5 | 129,832 | 0.53% | 3.6% | 1.15% | 95.7% |
| 6 | 558,919 | 0.79% | 1.34% | 0.03% | 97.83% |
| 7 | 114,077 | 0.03% | 1.47% | 0.02% | 98.47% |
| 8 | 2,954,230 | 0.17% | 2.46% | 2.4% | 97.32% |
| 9 | 2,753,884 | 6.5% | 1.57% | 0.18% | 91.7% |
| 10 | 1,309,791 | 8.11% | 1.2% | 0.002% | 90.67% |
| 11 | 107,251 | 7.6% | 2.53% | 0.002% | 89.85% |
| 12 | 325,471 | 0.65% | 2.34% | 0.007% | 96.99% |
| 13 | 1,925,149 | 2.01% | 1.65% | 0.06% | 96.26% |

Table (5): The labels distribution among each scenario

Each scenario has a unique combination of regular and malicious traffic. The CTU-13 has a specific malware executed for each scenario that uses several protocols and performs different actions. Table (6) shows the characteristics of the dataset for each botnet scenario. We can see the following acronyms in the table IRC (Internet Relay Chat), SPAM, CF (ClickFraud), PS (Port Scan), DDoS (Distributed Denial-of-Service), P2P (Peer-to-Peer), HTTP (Hypertext Transfer Protocol), and US (the traffic that complied and controlled by CTU-13 authors), where "Yes" means particular traffic is being used as part of the total traffic in that specific scenario, and "No" refers that traffic is not being used.

| Scenario | IRC | SPAM | CF | PS | DDoS | P2P | US | HTTP |
|---|---|---|---|---|---|---|---|---|
| 1 | Yes | Yes | Yes | No | No | No | No | No |
| 2 | Yes | Yes | Yes | No | No | No | No | No |
| 3 | Yes | No | No | Yes | No | No | Yes | No |
| 4 | Yes | No | No | No | Yes | No | Yes | No |
| 5 | No | Yes | No | Yes | No | No | No | Yes |
| 6 | No | No | No | Yes | No | No | No | No |
| 7 | No | No | No | No | No | No | No | Yes |
| 8 | No | No | No | Yes | No | No | No | No |
| 9 | Yes | Yes | Yes | Yes | No | No | No | No |
| 10 | Yes | No | No | No | Yes | No | Yes | No |
| 11 | Yes | No | No | No | Yes | No | Yes | No |
| 12 | No | No | No | No | Yes | No | No | No |
| 13 | No | Yes | No | Yes | No | No | No | Yes |

Table (6): The characteristics of each scenario in the dataset

After discarding unnecessary data, we have experimented with four scenarios, scenario #5 because it has the highest Normal Flows compared to the total flows, scenario #8 because it has the highest Command and Control (C&C) flows compared to the total flows, scenario #9 because it has the traffic combination that we are looking for, scenario #10 because it has the highest Bonet flows compared to the total flows. Table (7) represents the experiment and compares Precision, Recall, and F1-score for the selected datasets. The width window is set to 3.15 minutes, and the stride is set to 2.15 minutes. These values have been selected from Table (3) after running multiple experiments.

| Scenario | Precision | Recall | F1-score |
|---|---|---|---|
| 5 | 0.353 | 0.640 | 0.421 |
| 8 | 0.212 | 0.174 | 0.190 |
| 9 | 0.739 | 0.877 | 0.801 |
| 10 | 0.575 | 0.607 | 0.589 |

Table (7): Precision, Recall, and F1-score of different scenarios from the CTU-13 dataset

We can observe that scenario #9 has the best values of Precision, Recall, and F1-score, and in order to understand why we got such values, we have tested the highest values from Botnet Flows, Normal, and C&C Flows. We confirmed that our model works the best to detect Botnet Flows. As we can see in scenario #10, we have Botnet flows as the most significant portion of the data. It shows promising results compared to scenario #8, where C&C is the most significant portion of the data compared to scenario #5, with Normal high flows as the most significant portion of the data.

**5.4: Visualisation of the experiment**

In the experiments, in Figure (2), we have calculated Precision values for different 15 experiments with unique values of Width Window and Stride. The highest value was obtained in experiment number 6, which equals "0.745", where Width equals 3 minutes, and Stride equals 30 seconds. The lowest value has been obtained in experiment number 14 that equals "0.591", where Width equals 1 minute, and Stride equals 1 minute.

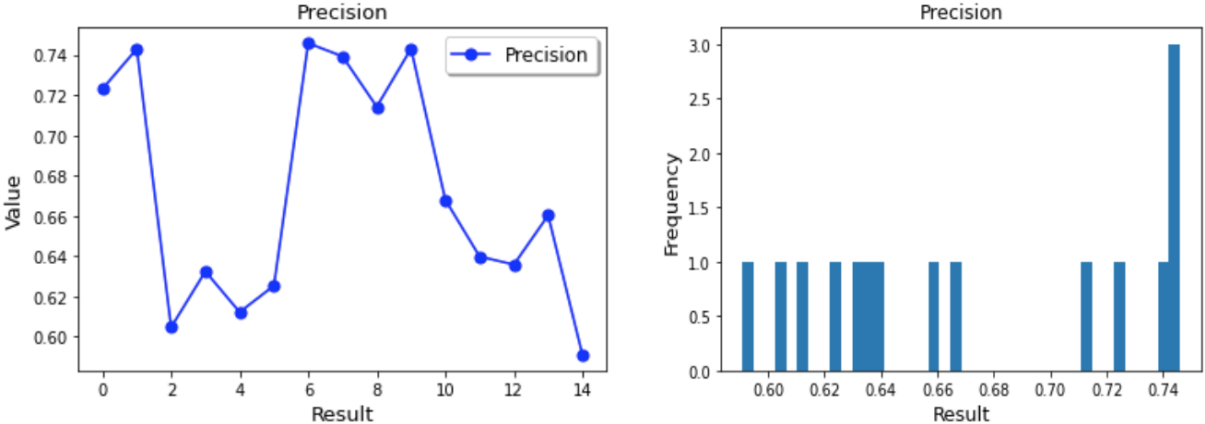

Figure (2): Precision values and histogram plot

We have calculated the histogram statistics to understand the occurrence of each value range to determine the reputation and the relation between the values. Histogram plot will make reading the data more straightforward. For example, suppose we want to understand how many precisions are above 0.70, then look at the plot. In that case, we can easily count six values, or how many precision values are below 0.60. From the histogram plot, we can say that only one value is below 0.60 in the experiments.

In Figure (3), we can see Recall values that have been calculated in the experiments. We can see that the highest value in experiment number 10 is "0.924", where Width equals 2.25 minutes, and Stride equals 15 seconds. The lowest value has been obtained in experiment number 8 that equals "0.787", where Width equals 7 minutes, and Stride equals 3 minutes. The histogram plot shows that we have 11 above 0.90 and only one value below 0.80 in the experiments.

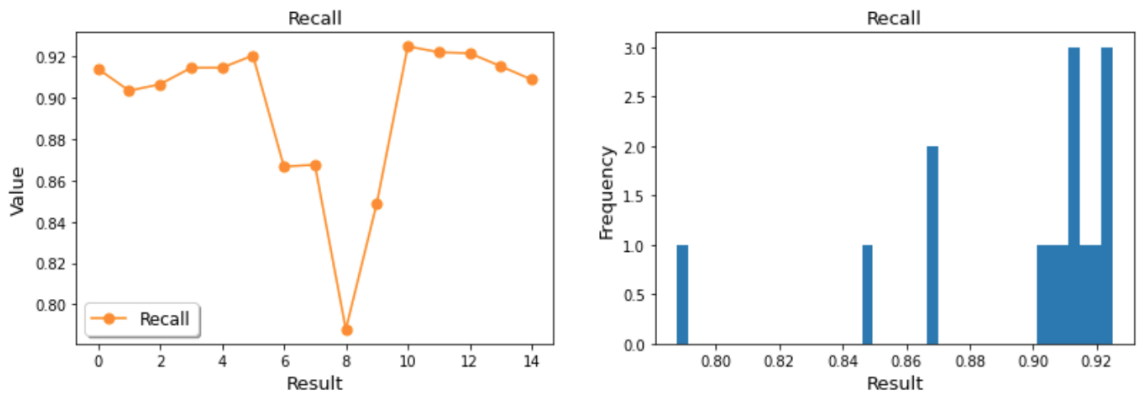

Figure (3): Recall values and histogram plot

In Figure (4), we can see F1-score values. The highest value has been obtained in experiment number 1 that equals "0.815", where Width equals 1.5 minutes, and Stride equals 15 seconds. The lowest value has been obtained in experiment number 14 that equals "0.716", where Width equals 1 minute, and Stride equals 1 minute. The histogram plot shows that we have three above 0.80 and only one value below 0.72 in the experiments.

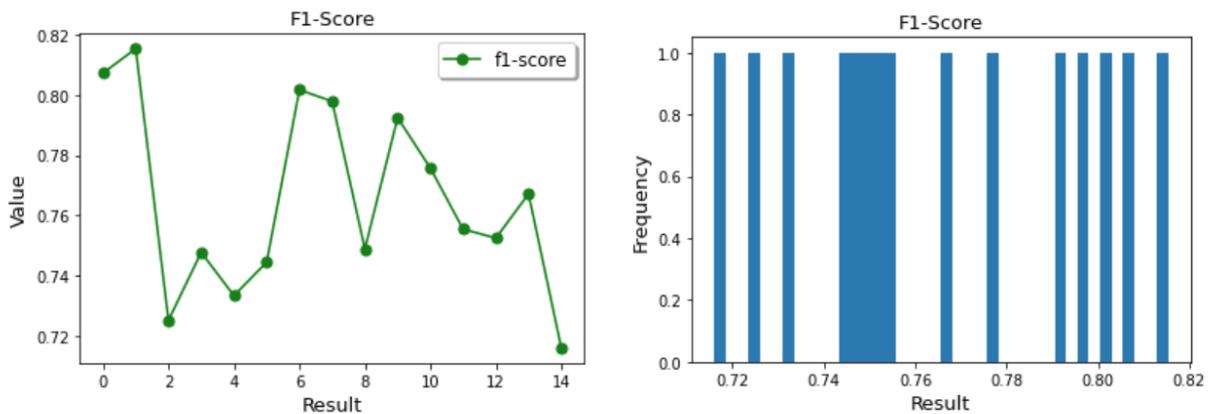

Figure (4): F1-Score values and histogram plot

The performance of our classifier is measured using the confusion matrix terminology. In Figure (5), our classifier has successfully predicted 1,168,470 True Negative and 1,617 True Positive, while the same classifier has misclassified 590 False Positive, and we got only 164 False Negative predictions.

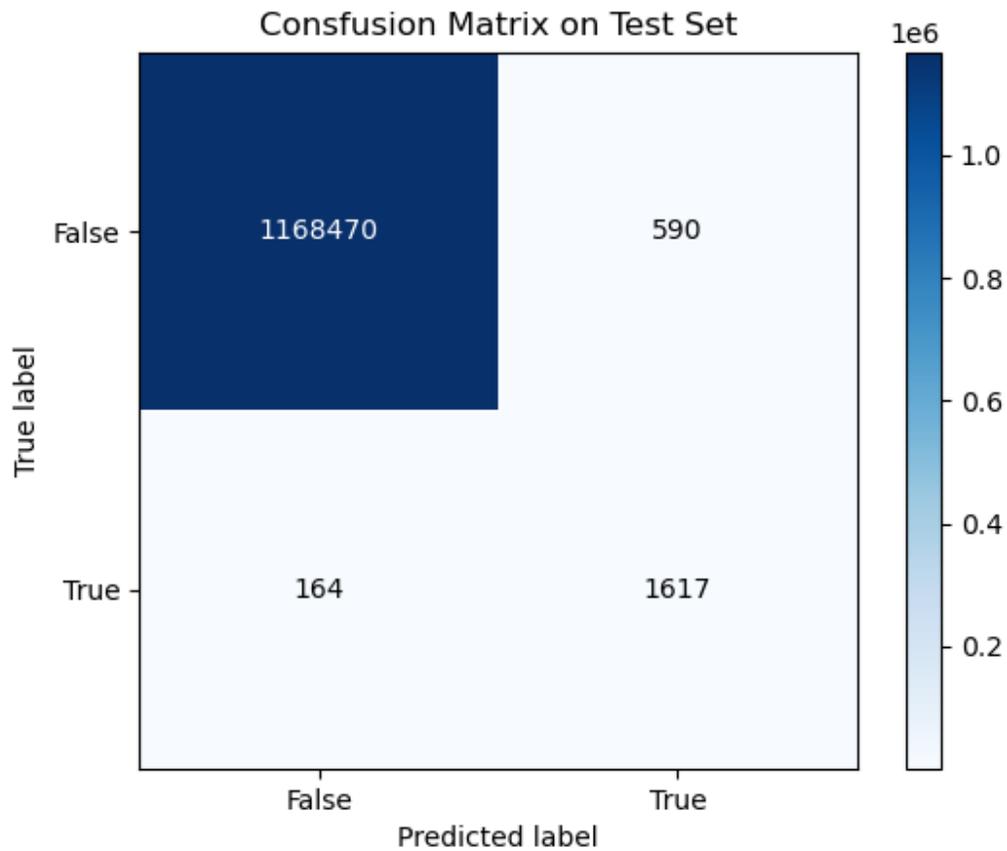

Figure (5): Scenario #9 Width is 90 seconds and Stride is 15 seconds

We tested a new classifier on the same dataset, scenario #9, by tuning the Width window to 189 seconds and the Stride window to 129 seconds. Our classifier has successfully predicted 276,989 True Negative and 205 True Positive, while the same classifier has misclassified 65 False Positive, and we got only 26 False Negative predictions.

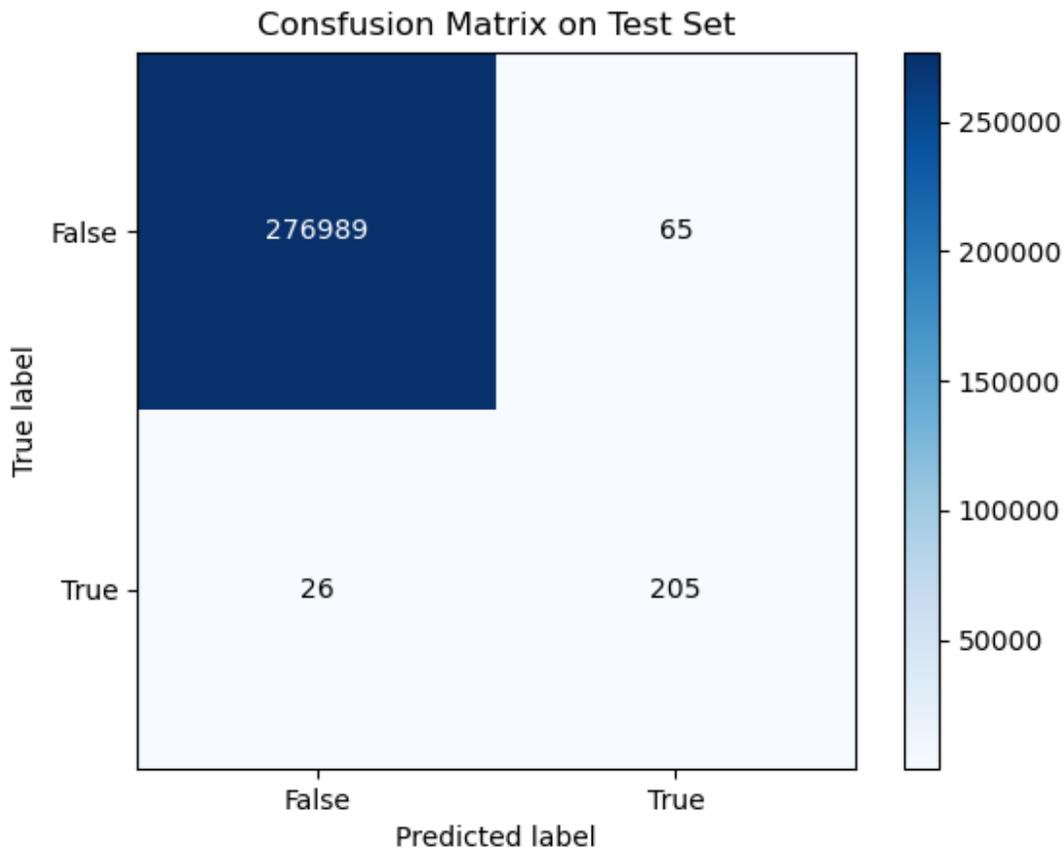

Figure (6): Scenario #9 Width is 189 seconds and Stride is 129 seconds

Using the same 189 seconds and 129 seconds in Width and Stride, respectively, we have tested other scenarios to confirm that our classifier works the best for Bot traffic detection. Scenario #5 contains a total of 129,832 flows that are being distributed as 0.53% (695 flows) of Botnet flows, 3.6% (4,679 flows) of Normal flows, 1.15% (206 flows) of Command&Control flows, and 95.7% (124,252 flows) of Background flows. The classifier detected only one True Positive and 14,951 True Negative. The classifier has detected 10 False Positive predictions with no False Negative. These experiments mimic real-world traffic activities where we can not predict the percentage of Bot, Malware, or Normal traffic. See Figure (7).

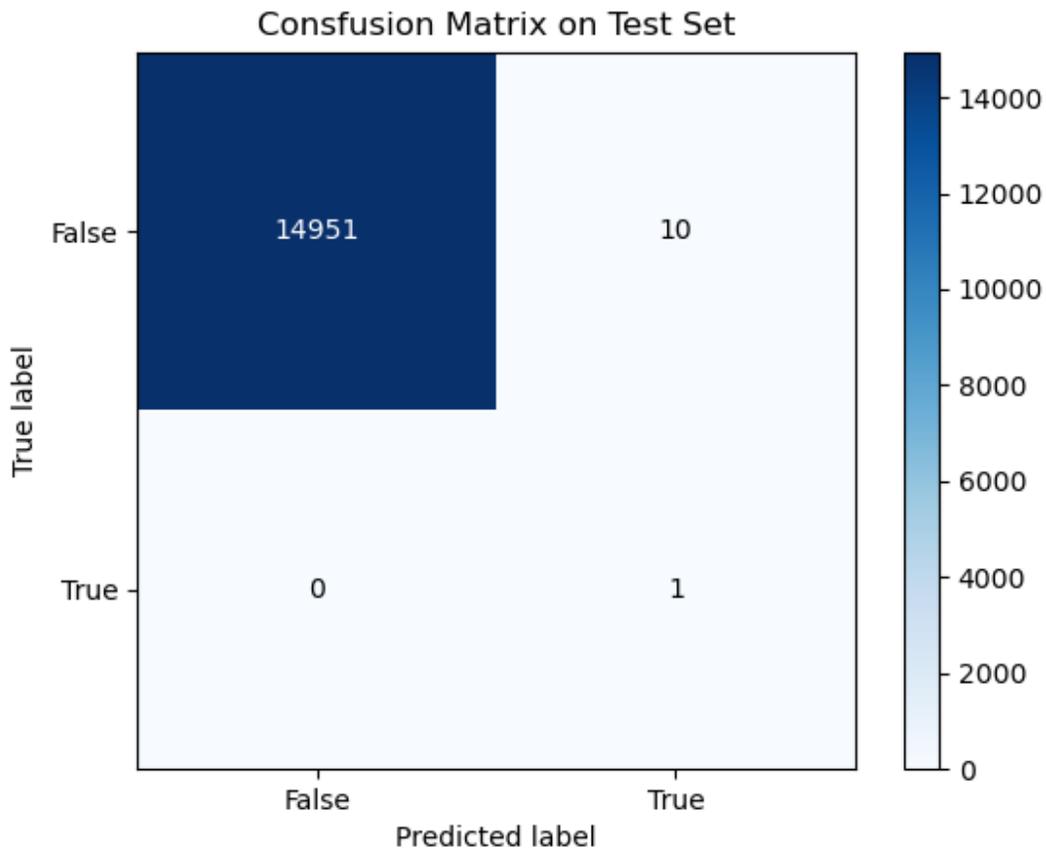

Figure (7): Scenario #5 Width is 189 seconds and Stride is 129 seconds

Our classifier failed to predict flows when there are very few Botnet flows in total flows, and there are high Command&Control flows and Normal flows. In Figure (8), we used scenario #8 that contains a total of 2,954,230 flows that are being distributed as 0.17% (5,052 flows) of Botnet flows, 2.46% (72,822 flows) of Normal flows, 2.4% (70,901 flows) of Command&Control (C&C) flows, and 97.32% (2,875282 flows) of Background flows. The classifier successfully predicted 387,943 True Negative and only 32 True Positive, while the classifier misclassified 118 False Positive and 155 False Negative. That shows that such a classifier can not be used to detect C&C flows.

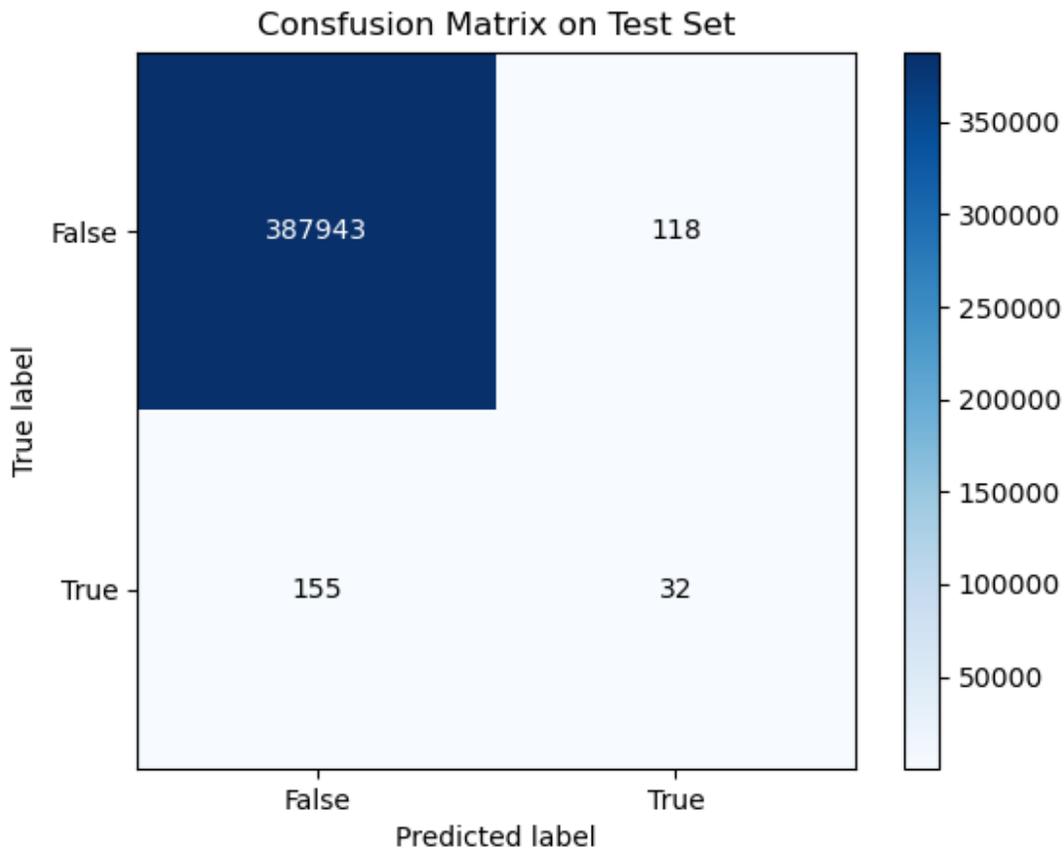

Figure (8): Scenario #8 Width is 189 seconds and Stride is 129 seconds

## 6) Conclusion

Processing big data is always challenging, and here analyses are building a reference for data processing of Logistic Regression Supervised Machine Learning algorithm. We observed that best values for recall come from Width= 135 seconds and Stride=15 seconds, and for precision from Width=180 seconds and Stride=30 seconds. We can test the same parameters and code in Table (3) on a different dataset in future work. Also, it is possible to change the dataset's features and run the experiment using different Machine Learning (ML) algorithms.